\documentclass[pra, twocolumn, floatfix, nofootinbib]{revtex4}
\usepackage{graphicx}
\usepackage{color}
\usepackage{amsmath, amsfonts, amssymb, bm}

\begin{document}
\title{Resonantly enhanced interatomic Coulombic electron capture\\ in a system of three atoms}
\author{S Remme}
\author{A B Voitkiv}
\author{C M\"uller}
\affiliation{Institut f\"ur Theoretische Physik I, Heinrich-Heine-Universit\"at D\"usseldorf, Universit\"atsstra{\ss}e 1, 40225 D\"usseldorf, Germany}
\date{\today}
\begin{abstract}
In interatomic Coulombic electron capture, the capture of a free electron at
an atomic center is accompanied by the radiationless transfer of the excess 
energy to a neighboring atom of different species, leading to ionization of the latter.
We show that this interatomic process can be strongly enhanced by the presence of an 
additional third atom, provided the energy of the free-bound capture transition 
in the first atom is resonant to a dipole-allowed excitation energy in this 
assisting atom. The relation of the resonantly enhanced three-center electron capture
with other processes is discussed, and its dependencies on the incident electron energy
and the spatial geometry of the triatomic system are illustrated.
\end{abstract}

\maketitle

\section{Introduction}
In electron capture processes, an initially free electron is captured 
by an atom or ion into a bound state. These elementary reactions, 
representing the inverse of ionization processes, are of relevance
for various areas of science, ranging from atomic and molecular physics
over plasma and astro-physics to chemistry and biophysics 
\cite{Review-Recomb1,Review-Recomb2,Review-Recomb3}.

Capture into atomic centers, that are isolated in space, can proceed in
three different ways. (i) The incident electron may transition directly into
an empty bound atomic state, releasing the energy difference via photoemission.
In this radiative recombination, other atomic electrons are merely spectators.
(ii) For certain energies of the incident electron, the capture can be accompanied 
by the simultaneous excitation of a bound atomic electron, leading to the formation 
of an autoionizing state, that afterwards stabilizes by photoemission. Due to
its resonant nature, the cross section for this dielectronic recombination 
can be strongly enhanced. (iii) If another free electron is available to absorb 
the excess energy, capture may also proceed via three-body recombination.

When a free electron is captured into an atomic center, that is not isolated 
in space but rather close to another atomic center, additional channels exist 
that rely on two-center electron-electron correlations. Corresponding interatomic 
processes have been under active scrutiny in recent years 
\cite{ICD-Review1, ICD-Review2, ICD-Review3}. As a two-center analogue of dielectronic 
recombination, an incident electron with suitable energy can be captured by an atom, 
transfering the excess energy to a neighboring atom of different species that is 
resonantly excited and subsequently deexcites via photoemission. This two-center 
dielectronic recombination (2CDR) can largely dominate over the single-center radiative 
recombination of the electron with the first atom at interatomic distances up to few 
nanometers \cite{2CDR-PRL,2CDR-PRA2010,2CDR-PRA2018}. The process may also occur in slow 
atomic collisions \cite{2CDR-coll,2CDR-coll-Hbar}. The inverse of 2CDR is two-center 
resonant photoionization \cite{2CDR-PRA2010,2CPI}; it has been observed experimentally in 
noble gas dimers \cite{2CPIexp} and clusters \cite{Hergenhahn}.

When the energy difference of the capturing transition at the first atomic center 
exceeds the ionization energy of a neighboring atom, yet another two-center electron 
capture process may occur. In interatomic Coulombic electron capture (ICEC) an 
incident electron is captured by an atom or ion, transferring the energy difference 
radiationlessly to a neigbouring atom, which is prompted this way to emit an electron 
\cite{ICEC-JPB, ICEC-PRA, ICEC-Sisourat}\footnote{The second step of ICEC is called 
interatomic Auger decay \cite{Matthew} or interatomic Coulombic decay (ICD) \cite{ICD}; 
it has been studied extensively in a variety of systems during the last two decades 
\cite{ICD-Review1,ICD-Review2,ICD-Review3}.}. 
ICEC thus induces effectively a charge exchange between the centers. It has also been 
studied in slow atomic collisions \cite{ICEC-coll} and in condensed-matter systems of 
two quantum dots \cite{ICEC-dots1,ICEC-dots2,ICEC-dots3}. As biophysical example,
electron attachment to a proton via ICEC in water was investigated \cite{ICEC-water}. 
ICEC can be considered as related to three-body recombination, with the excess energy being 
transferred, though, to an intially bound atomic electron rather than another free electron.

In its original form, ICEC is a nonresonant process. However,
for certain incident energies it allows for a resonant variant, when the electron 
capture at the first atomic center is accompanied by the simultaneous excitation 
of an atomic electron at the same center. This autoionizing state may then relax
either by photoemission, this way completing an ordinary single-center dielectronic 
recombination, or by energy transfer to and consequent ionization of a neighbouring 
atom. In the second case, the process is called resonant ICEC. It has been investigated in 
systems of two quantum dots \cite{res-ICEC1,res-ICEC2}. 

\begin{figure}[t]
\begin{center}
\includegraphics[width=0.49\textwidth]{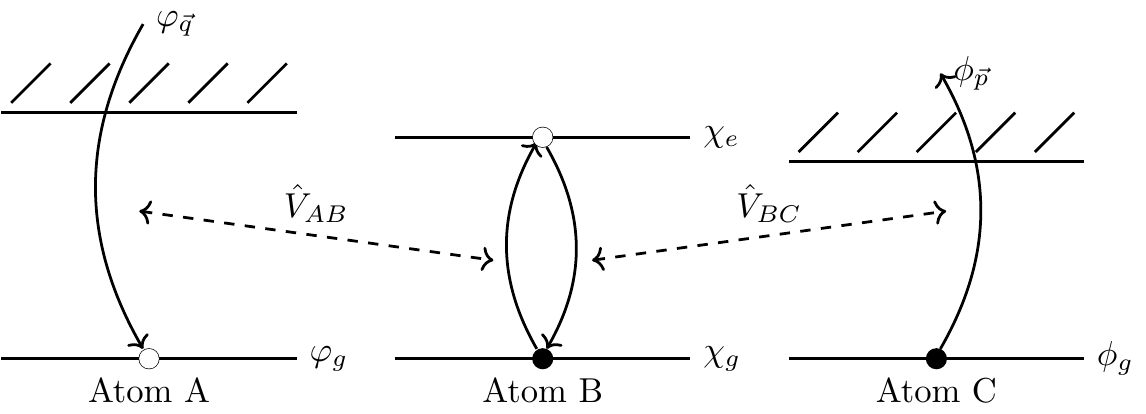}
\caption{Scheme of resonant electron capture in a triatomic system (three-center 
rICEC). In the first step, an incident electron is captured at center $A$, with the 
excess energy being transfered via interatomic electron-electron correlations 
to atom $B$, which is resonantly excited. In the second step, atom $B$ deexcites 
and transfers the excitation energy to atom $C$ that is consequently ionized.}
\label{fig:scheme}
\end{center}
\end{figure}

In the present paper, we consider another version of resonant ICEC that involves
three participating atoms: a free electron is captured at an atomic center $A$, 
transferring the energy difference resonantly to a neigbouring atom $B$ that is 
excited and, afterwards, transfers the excitation energy again radiationlessly 
to a third atom $C$, which is consequently ionized (see Fig.~\ref{fig:scheme}). The 
process may be termed three-center resonant electron capture (three-center rICEC). We show that, 
under suitable conditions, the participation of the 'catalyzer' atom $B$ can strongly 
enhance the electron capture as compared with the direct ICEC between atoms $A$ and $C$.

We note that the usual (nonresonant) ICEC and three-center rICEC are related in a similar 
way as single-center radiative recombination and 2CDR: by adding another atom to an 
$N$-center system, the resulting $(N+1)$-center system possesses additional 
electronic resonances which may cause resonant enhancements over the corresponding 
nonresonant processes in the $N$-center systems. It is worth mentioning in this context, 
that ICD has recently been shown to benefit from the presence of additional 'bridge' atoms, 
as well \cite{super-ICD}. In this case, however, virtual intermediate states of the bridge 
atoms play the crucial role, whereas real resonances are populated in three-center rICEC.
We furthermore refer to Refs.~\cite{Exp1,Exp2,Exp3,Exp4} for recent experimental work on 
ICD-related processes that are induced by electron impact.

Our paper is organized as follows. In Sec.~II we present our theoretical approach to 
three-center rICEC and discuss its relation with other capture processes. In Sec.~III 
we apply our formalism to a simple triatomic model system composed of hydrogen, helium 
and lithium and illustrate the dependencies of rICEC on the incident electron energy 
and the spatial geometry of the system. Conclusions are given in Sec.~IV. 
Atomic units (a.u.) are used throughout unless explicitly stated otherwise.

\section{Theory of three-center resonant electron capture}
\subsection{General considerations}
\begin{figure}[t]
\begin{center}
\includegraphics[width=0.49\textwidth]{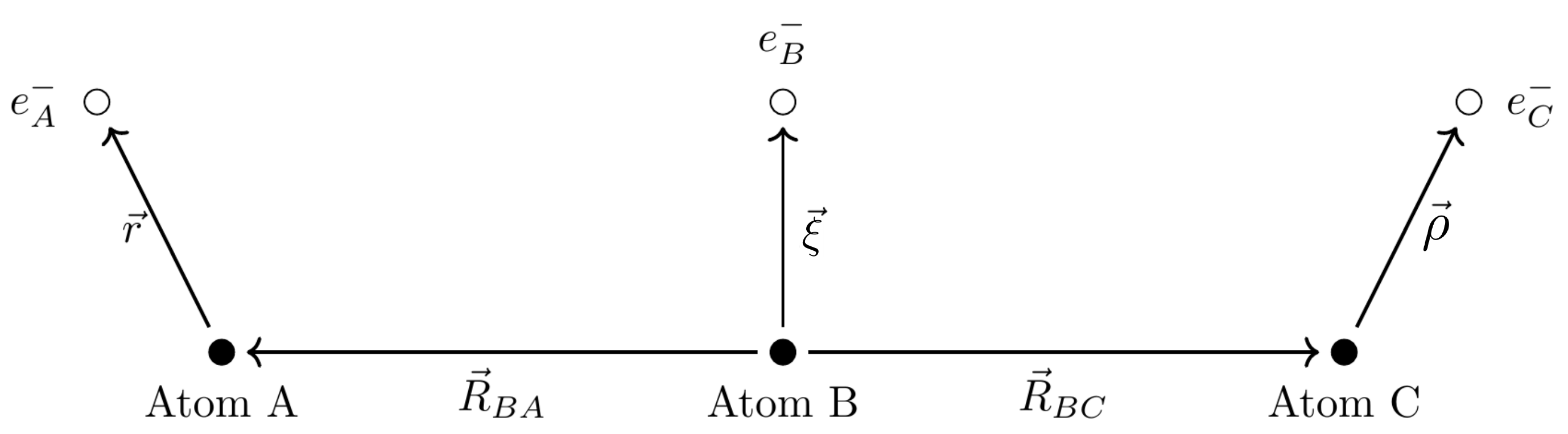}
\caption{Spatial coordinates of the triatomic system.}
\label{fig:coordinates}
\end{center}
\end{figure}

We consider a system consisting of three atoms $A$, $B$ and $C$, separated 
by sufficiently large distances such that their atomic individuality is basically preserved. 
Assuming the atoms to be at rest, we take the position of the nucleus of atom $B$  
as the origin and denote the coordinates of the nucleus of atom $A$ by ${\bf R}_{BA}$ 
and the nucleus of atom $C$ by ${\bf R}_{BC}$ (see Fig.~\ref{fig:coordinates}).
Accordingly, ${\bf R}_{AC} = {\bf R}_{BC}-{\bf R}_{BA}$ is the separation
between the atomic centers $A$ and $C$. 

The coordinates of the electron incident on atom $A$ are ${\bf r}_A = {\bf R}_{BA}
+ \bf{r}$, such that ${\bf r}$ denotes the position vector with respect to the
nucleus of atom $A$. Similarly are ${\bf r}_C = {\bf R}_{BC}+\boldsymbol{\rho}$
the coordinates of the electron in atom $C$. In this section we assume for 
definiteness (and simplicity) that in atom $B$ there is one active electron 
with position vector $\boldsymbol{\xi}$. Generalization to more than one electron
is straightforward; an example is discussed in Sec.~III. Besides, atom $B$ is supposed to have an 
excited state $\chi_e$ reachable from the ground state $\chi_g$ by a dipole-allowed transition. 

The total Hamiltonian describing the triatomic system reads 
\begin{eqnarray} 
H =  \hat{H}_0 + \hat{V}_{AB} + \hat{V}_{BC} + \hat{V}_{AC},  
\label{hamiltonian}  
\end{eqnarray} 
where $ \hat{H}_0 $ is the sum of the Hamiltonians for the noninteracting 
atoms and $\hat{V}_{AB}$, $\hat{V}_{BC}$ and $\hat{V}_{AC}$
denote the mutual interactions between the respective pair of atoms.
For electrons undergoing electric dipole transitions and $R_{BA}\gg 1$\,a.u., 
the interatomic interaction reads 
\begin{eqnarray} 
\hat{V}_{AB}({\bf r},\boldsymbol{\xi};{\bf R}_{BA}) = \frac{{\bf r}\cdot \boldsymbol{\xi}}{R_{BA}^3} 
- \frac{ 3 ({\bf r}\cdot{\bf R}_{BA})(\boldsymbol{\xi}\cdot{\bf R}_{BA})}{R_{BA}^5}\ .
\label{V_AB} 
\end{eqnarray}
Corresponding expressions hold {\it mutatis mutandis} for 
$\hat{V}_{BC}(\boldsymbol{\rho},\boldsymbol{\xi};{\bf R}_{BC})$ and
$\hat{V}_{AC}({\bf r},\boldsymbol{\rho};{\bf R}_{AC})$. Since the internuclear distances
are assumed to be rather large, the interatomic interactions may be treated perturbatively.

In the process of three-center rICEC one has essentially three different basic two-electron configurations, which are schematically illustrated in Fig.~\ref{fig:scheme}:\\
(I) $\Psi_{{\bf q},g,g} = \varphi_{\bf q}({\bf r})\chi_{g}(\boldsymbol{\xi})\phi_g(\boldsymbol{\rho})$ with total energy $E_{{\bf q},g,g} = \varepsilon_q + e_0 + \epsilon_0$, where a free electron with asymptotic momentum ${\bf q}$ and energy $\varepsilon_q=\frac{q^2}{2}$ is incident on atom $A$, while atoms $B$ and $C$ are in their ground states $\chi_g$ and $\phi_g$;
(II) $\Psi_{g,e,g} = \varphi_{g}({\bf r}) \chi_e(\boldsymbol{\xi}) \phi_g(\boldsymbol{\rho})$ with total energy $E_{g,e,g} = \varepsilon_0 + e_1 + \epsilon_0$, in which atoms $A$ and $C$ are is their ground states, while atom $B$ is in the excited state $\chi_e$;
(III) $\Psi_{g,g,{\bf p}} = \varphi_g({\bf r}) \chi_g(\boldsymbol{\xi}) \phi_{{\bf p}}(\boldsymbol{\rho})$ with total energy $E_{g,g,{\bf p}} = \varepsilon_0 + e_0 + \epsilon_p$, where atoms $A$ and $B$ are in their ground states, while from atom $C$ an electron has been emitted into the continuum with asymptotic momentum ${\bf p}$ and energy $\epsilon_p=\frac{p^2}{2}$. 

Within the second order of time-dependent perturbation theory, the probability amplitude for three-center rICEC can be written as
\begin{eqnarray}
S_{\bf p}^{(r)}\! &=&\! -\int\limits_{-\infty}^{\infty} {\rm d}t\, \langle \Psi_{g,g,{\bf p}}|\hat{V}_{BC}| \Psi_{g,e,g} \rangle\, e^{-i(E_{g,e,g}-E_{g,g,{\bf p}})t}\nonumber\\
& &\!\!\!\!\! \times\!\! \int\limits_{-\infty}^{t} {\rm d}t'\, \langle \Psi_{g,e,g}|\hat{V}_{AB}| \Psi_{{\bf q},g,g} \rangle\, {\rm e}^{-i(E_{{\bf q},g,g}-E_{g,e,g})t'}
\label{S1}
\end{eqnarray}
When there are several energetically degenerate excited states in atom $B$ (distinguished, e.g., by their magnetic quantum number) the transition amplitude is to be amended by a coherent sum over these intermediate states. 

By performing the inner time integral, we obtain
\begin{eqnarray*} 
S_{\bf p}^{(r)} &=& -i\int_{-\infty}^{\infty} dt\, \langle \chi_g \phi_{{\bf p}}|\hat{V}_{BC}| \chi_{e}\phi_g \rangle\nonumber \\
& & \times\,\frac{\langle\varphi_g\chi_e|\hat{V}_{AB}|\varphi_{\bf q}\chi_g \rangle\,{\rm e}^{-i(\varepsilon_q + \epsilon_0 - \varepsilon_0 - \epsilon_p)t}}{\varepsilon_q + e_0 - \varepsilon_0 -e_1 + \frac{i}{2}\Gamma}\ .
\label{S2}
\end{eqnarray*}
Here we have inserted the total width $\Gamma=\Gamma_{\rm rad}^{(B)}+\Gamma_{\rm ICD}^{(BA)} + \Gamma_{\rm ICD}^{(BC)}$ of the excited state $\chi_e$ in atom $B$. It accounts for the finite lifetime of this state and consists of the radiative width $\Gamma_{\rm rad}^{(B)}$ and the ICD widths $\Gamma_{\rm ICD}^{(BA)}$ and $\Gamma_{\rm ICD}^{(BC)}$ associated with the interatomic Coulombic decay of $\chi_e$ involving ionization of either atom $A$ or atom $C$.
These widths are given by
\begin{eqnarray*}
\Gamma_{\rm rad}^{(B)}\!&=&\!\frac{4(e_1-e_0)^3}{3c^3}\big| \langle \chi_e|\boldsymbol{\xi}|\chi_g\rangle \big|^2\ , \\
\Gamma_{\rm ICD}^{(BA)}({\bf R}_{BA})\! &=&\! \frac{q}{(2\pi)^2}\int{\rm d}\Omega_{{\bf q}'} \big|\langle\phi_{{\bf q}'}\chi_g|\hat{V}_{AB}|\phi_g\chi_e\rangle\big|^2
\end{eqnarray*}
where the integral is taken over the emission angles of the ICD electron that is reejected from atom $A$;
a corresponding expression holds for $\Gamma_{\rm ICD}^{(BC)}({\bf R}_{BC})$.

Taking also the outer time integral, we arrive at
\begin{eqnarray} 
S^{(r)}_{\bf p}\! &=&\! \frac{\langle \chi_g \phi_{{\bf p}}|\hat{V}_{BC}| \chi_{e}\phi_g \rangle\, \langle\varphi_0\chi_e|\hat{V}_{AB}|\varphi_{\bf q}\chi_g \rangle}{\Delta + \frac{i}{2}\Gamma} \nonumber \\
& & \times\, 2\pi\delta(\varepsilon_q + \epsilon_0 - \varepsilon_0 - \epsilon_p)\ ,
\label{S3}
\end{eqnarray}
where the detuning from the resonance $\Delta = \varepsilon_p + e_0 - \varepsilon_0 - e_1$ has been introduced.
The delta function in Eq.~\eqref{S3} displays the energy conservation in the process. In this relation, the energies of atom $B$ have dropped out, in accordance with its role as catalyzer.

From the transition amplitude we can obtain the cross section in the usual way by taking the absolute square, integrating it over the momentum of the outgoing electron, and dividing it by the interaction time $\tau$ and the incident electron flux $j=q$, that is
\begin{eqnarray} 
\sigma^{(r)} = \int \frac{{\rm d}^3p}{(2\pi)^3 j\tau}\,\big|S_{\bf p}^{(r)}\big|^2\ .
\label{CS}
\end{eqnarray}
Note that the continuum states in our calculations are normalized to a quantization volume of unity. 

\subsection{Relations to other processes}
In the general case of arbitrary interatomic separation vectors ${\bf R}_{BA}$ und ${\bf R}_{BC}$, the transition amplitude \eqref{S3} possesses a rather involved dependence on the spatial geometry of the system (see Sec.~III.B). In order to reveal some basic properties of three-center rICEC, we consider in this subsection the special case when the incident momentum ${\bf q}$ and the interatomic separation vectors ${\bf R}_{BA}$ and ${\bf R}_{BC}$ all lie on the same axis, that is chosen along $z$. 

In this situation, the expression in Eq.~\eqref{S3} simplifies to
\begin{eqnarray} 
S_{\bf p}^{(r)} &=& \frac{\langle \phi_{{\bf p}}|\rho_z|\phi_g \rangle\,
\big|\langle \chi_e |\xi_z| \chi_{g} \rangle\big|^2\, \langle\varphi_g|z|\varphi_{\bf q}\rangle}{\big(\Delta + \frac{i}{2}\Gamma\big) R_{BA}^3 R_{BC}^3} \nonumber \\
& & \times\, 8\pi\delta(\varepsilon_q + \epsilon_0 - \varepsilon_0 - \epsilon_p)\ .
\label{S4}
\end{eqnarray}
By using the corresponding amplitude for direct ICEC between atoms $A$ and $C$ 
\begin{eqnarray}
S^{(d)}_{\bf p}\! &=&\! -i\int\limits_{-\infty}^{\infty} {\rm d}t\, \langle \varphi_g\phi_{\bf p}|\hat{V}_{AC}|\varphi_{\bf q}\phi_g \rangle\, {\rm e}^{-i(\varepsilon_q + \epsilon_0 - \varepsilon_0 - \epsilon_{q})t}\nonumber\\
&=&\! \frac{4\pi i}{R_{AC}^3} \langle\phi_{\bf p}|\rho_z| \phi_g \rangle \langle \varphi_g|z|\varphi_{\bf q}\rangle\,\delta(\varepsilon_q + \epsilon_0 - \varepsilon_0 - \epsilon_p)
\label{S-ICEC}
\end{eqnarray}
the expression in Eq.~\eqref{S4} can be cast in the form
\begin{eqnarray} 
S^{(r)}_{\bf p} = -\frac{2i R_{AC}^3\,
\big|\langle \chi_e |\xi_z| \chi_{g} \rangle\big|^2}{R_{BA}^3 R_{BC}^3\big(\Delta + \frac{i}{2}\Gamma\big)} \, S^{(d)}_{\bf p}\ .
\label{S5}
\end{eqnarray}
When the resonance condition is met ($\Delta=0$) and the interatomic distances are sufficiently large, so that the dominant contribution to the total with $\Gamma$ comes from the radiative width $\Gamma_{\rm rad}^{(B)}$, this relation becomes 
$ S^{(r)}_{\bf p} = -3 [cR_{AC}/(\omega_{ge} R_{BA}R_{BC})]^3 S^{(d)}_{\bf p}$,
which -- for the considered geometry -- implies
\begin{eqnarray} 
\sigma^{(r)}\, =\, 9 \left(\frac{cR_{AC}}{\omega_{ge} R_{BA}R_{BC}}\right)^{\!6} \sigma^{(d)}
\label{CS-2}
\end{eqnarray}
Here we have inserted the formula for the radiative width given above with $\omega_{ge}=e_1-e_0$. 

Equation\,\eqref{CS-2} shows that the cross section for three-center rICEC can be largely enhanced as compared with the direct (nonresonant) ICEC where the catalyzer atom $B$ is not involved. Assuming that $R_{BA}\approx R_{BC}\approx R_{AC}$, the enhancement factor is of order $9[c/(\omega_{ge} R)]^6\gg 1$, where $R$ can be taken as one of the interatomic distances. For example, assuming $\omega_{ge}\approx 20$\,eV and $R\approx 10$\,\AA, an enormous enhancement by seven orders of magnitude results. In a linear configuration $A$-$B$-$C$ with $R_{BA}\approx R_{BC}\approx \frac{1}{2}R_{AC}$ like in Fig.~\ref{fig:coordinates}, the enhancement is even further amplified by an additional factor $\approx 2^6$.

Since three-center rICEC and direct ICEC lead to the same final state, the corresponding transition amplitudes \eqref{S3} and \eqref{S-ICEC} are generally subject to quantum interference. This effect will be illustrated in Sec.~III.A. We note that, for parameters where rICEC strongly dominates, the interference is of minor importance and may be neglected.

The cross section of three-center rICEC can also be related to various single-center quantities. To this end, we note that the matrix elements for free $\leftrightarrow$ bound transitions in Eq.~\eqref{S4} are contained in the probability amplitudes for the single-center processes of radiative recombination of a free electron with atom $A$ and photoionization of atom $C$, respectively. Accordingly, one can derive the following compact expression for the cross section of three-center rICEC:
\begin{equation}
\sigma^{(r)} = \frac{9}{4}\left(\frac{c}{\omega_{ge}R_{BA}}\right)^{\!6}\,\frac{\Gamma_{\rm rad}^{(B)}\,\Gamma_{\rm ICD}^{(BC)}}{\Delta^2+\frac{1}{4}\Gamma^2}\ \sigma_{\rm RR}^{(A)}
\label{CS-3}
\end{equation}
where $\sigma_{\rm RR}^{(A)}$ denotes the cross section for single-center radiative recombination of the incident electron into atom $A$ and the relation $\Gamma_{\rm ICD}^{(BC)}=\frac{3c^4}{2\pi\omega_{ge}^4 R_{BC}^6}\,\Gamma_{\rm rad}^{(B)}\,\sigma_{\rm PI}^{(C)}$ has been used (see, e.g., \cite{ICD-formula}), with the single-center photoionization cross section $\sigma_{\rm PI}^{(C)}$ of atom $C$. Equation~\eqref{CS-3} shows that three-center rICEC can also largely dominate over the single-center process of radiative recombination. Moreover, its factorized structure allows for an intuitive interpretation: the cross section $\sigma_{\rm RR}^{(A)}$ represents a measure for the probability that the incident electron is captured by atom $A$; the resonant enhancement factor $[c/(\omega_{ge}R_{BA})]^6$ together with $\Gamma_{\rm rad}^{(B)}$ stand for the resonant excitation of atom $B$; and the ICD width $\Gamma_{\rm ICD}^{(BC)}$ describes the final step, where atom $B$ deexcites with simultaneous electron emission from atom $C$.\footnote{Formulas corresponding to Eq.~\eqref{CS-2} for direct ICEC and resonant ICEC in two-center systems have been obtained in Refs.~\cite{ICEC-JPB,ICEC-PRA} and \cite{res-ICEC1}, respectively.}

An expression similar to Eq.~\eqref{CS-3}, with the ICD width $\Gamma_{\rm ICD}^{(BC)}$ being replaced by another factor of the radiative width $\Gamma_{\rm rad}^{(B)}$, describes the cross section of 2CDR involving atoms $A$ and $B$ \cite{2CDR-PRL,2CDR-PRA2010,2CDR-PRA2018}. The origin of this difference is clear: in three-center rICEC the excited state in atom $B$ relaxes via ICD with atom $C$, whereas in 2CDR it decays radiatively. Accordingly, three-center rICEC dominates over the competing process of 2CDR in a triatomic system when $\Gamma_{\rm ICD}^{(BC)} > \Gamma_{\rm rad}^{(B)}$. When three-center rICEC in a system $A$-$B$-$C$ is compared instead with 2CDR in a diatomic system $A$-$B$, the condition of dominance reads $\Gamma_{\rm ICD}^{(BC)}/\Gamma^2 > \Gamma_{\rm rad}^{(B)}/\Gamma_{\rm 2CDR}^2$, where $\Gamma_{\rm 2CDR} = \Gamma_{\rm rad}^{(B)} + \Gamma_{\rm ICD}^{(BA)} < \Gamma$ denotes the total width in the diatomic 2CDR case.

Concluding this section, the process of three-center rICEC has turned out to represent an additional channel for capture of an electron to center $A$ in a triatomic system $A$-$B$-$C$, that can strongly outperform by several orders of magnitude not only direct ICEC between $A$ and $C$, but also radiative recombination into an isolated center $A$ or 2CDR involving centers $A$ and $B$.

\section{Numerical Results and Discussion}
In this section, we illustrate characteristic properties of three-center rICEC by way of a concrete numerical example. To this end, we consider a triatomic model system that is composed of the three lightest elements hydrogen, helium and lithium. Symbolically, the process proceeds according to 
$$ e + {\rm H}^+ + {\rm He} + {\rm Li} \to {\rm H} + {\rm He}^* + {\rm Li} \to {\rm H} + {\rm He} + {\rm Li}^+ + e'\ ,$$
whereas the direct (nonresonant) ICEC is described by
$$ e + {\rm H}^+ + {\rm He} + {\rm Li} \to {\rm H} + {\rm He} + {\rm Li}^+ + e'\ ,$$
with the helium atom being a mere spectator. In both cases, the electron capture occurs into the $1s$ ground state of hydrogen, and the $2s$ electron is emitted from lithium. Note that in the considered system, two-center resonant ICEC between the atoms $A$ and $C$ is not possible.

In Sec.~II we developed our theory of three-center rICEC assuming one (active) electron per atom. This assumption is exactly met for hydrogen and to a good approximation in lithium as well. Accordingly, we describe the initial and final state of the captured electron by a Coulomb wave function \cite{LL} with asymptotic momentum ${\bf q}=q\,{\bf e}_z$ and a $1s$ wave function, both for a nuclear charge $Z_A=1$. For the states in lithium, we chose an effective nuclear charge $Z_C=1.259$ to match the binding energy $\epsilon_0\approx -5.39$\,eV \cite{NIST} of the valence electron and describe its state by a hydrogenic $2s$ wave function. The ejected electron is described by a Coulomb wave function for the same nuclear charge and with asymptotic momentum ${\bf p}$.

In helium, however, there are two equivalent electrons; we shall therefore work with properly symmetrized two-electron helium states. The ground and excited states accordingly read
\begin{eqnarray*}
\chi_g(\boldsymbol{\xi}_1,\boldsymbol{\xi}_2) \!&=&\! \alpha_{1s}(\boldsymbol{\xi}_1)
\alpha_{1s}(\boldsymbol{\xi}_2)\ ,\\
\chi_e^{(m)}(\boldsymbol{\xi}_1,\boldsymbol{\xi}_2) \!&=&\! \frac{1}{\sqrt{2}}\big[
\alpha_{2p_m}(\boldsymbol{\xi}_1)\alpha_{1s}(\boldsymbol{\xi}_2) + \alpha_{1s}(\boldsymbol{\xi}_1)
\alpha_{2p_m}(\boldsymbol{\xi}_2)\big]
\end{eqnarray*}
During three-center rICEC, the helium atom is assumed to be resonantly excited from the $(1s)^2$ ground state to a $1s2p_m$ excited state with magnetic quantum number $m\in\{0, \pm 1\}$. The $\alpha_{2p_m}$ and $\alpha_{1s}$ are taken as hydrogenic wave functions with effective nuclear charge $Z_B=1.435$ to match the transition energy of $\omega_{ge}\approx 21.22$\,eV \cite{NIST}. 
The interatomic interactions have to be amended, accordingly, in order to account for the two electrons in helium. They can be obtained from Eq.~\eqref{V_AB} as $\hat{V}_{AB}({\bf r},\boldsymbol{\xi}_1 +\boldsymbol{\xi}_2;{\bf R}_{BA})$ and $\hat{V}_{BC}(\boldsymbol{\rho},\boldsymbol{\xi}_1 + \boldsymbol{\xi}_2;{\bf R}_{BC})$. 

\subsection{Resonant enhancement and energy dependence}

We first illustrate the dependence of three-center rICEC on the incident electron energy. For definiteness, the atoms of our model system are assumed to form a linear configuration along the $z$ axis (see Fig.~\ref{fig:scheme}), with internuclear distances $R_{BA}=R_{BC}=10$\,a.u. In this case, solely the $1s2p_0$ excited state in helium can be populated.

\begin{figure}[t]
\begin{center}
\includegraphics[width=0.45\textwidth]{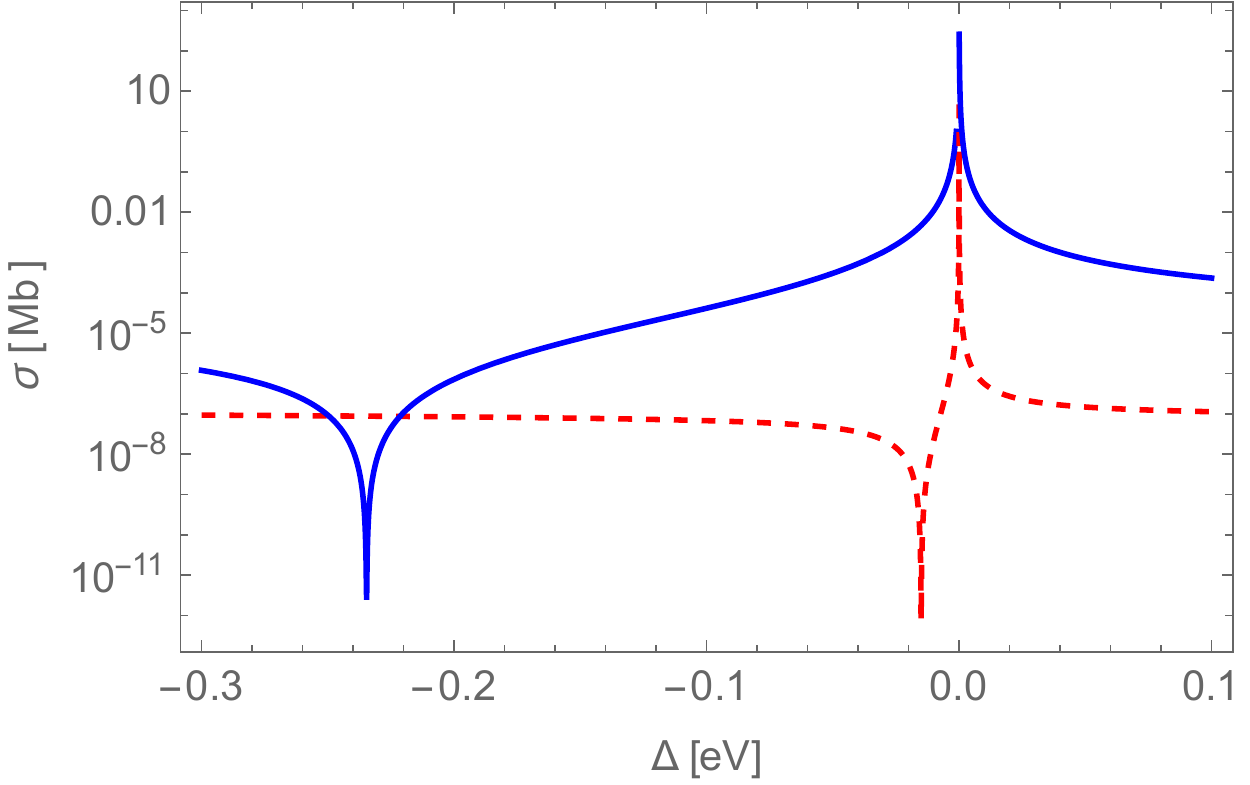}
\caption{Cross section for three-center rICEC in a H-He-Li system 
with $R_{BA}=R_{BC}=10$\,a.u. (blue solid curve) and $R_{BA}=R_{BC}=25$\,a.u. 
(red dashed curve), as function of the detuning of the incident electron 
energy from the resonant value. The interference with direct ICEC
between H and Li is included, leading to characteristic Fano profiles.}
\label{fig:energy}
\end{center}
\end{figure}

Figure~\ref{fig:energy} depicts the dependence of the cross section for three-center rICEC on the detuning $\Delta = \varepsilon_q - \varepsilon_0 - \omega_{ge}$. At exact resonance, where the incident electron energy amounts to $\varepsilon_q\approx 7.62$\,eV, the cross section is sharply peaked, reaching the value $\sigma^{(r)}\approx 2.6\times 10^{-16}$\,cm$^2$. For comparison, we note that the corresponding cross section for direct ICEC amounts to $\sigma^{(d)}\approx 2.2\times 10^{-23}$\,cm$^2$. 
In light of this huge difference, the quantum interference that both processes are subject to, is immaterial when the energy lies close to the resonance. Only at very large detuning ($\Delta\approx -0.24$\,eV), the interference effects become relevant in the form of a characteristic Fano minimum (see Refs.~\cite{2CDR-PRL,2CDR-PRA2010,2CDR-PRA2018,2CPI,ICEC-water} for similar interference structures). For comparison, Fig.~\ref{fig:energy} also displays the cross section for enlarged internuclear distances of $R_{BA}=R_{BC}=25$\,a.u. Here, the three-center rICEC cross section reaches a maximum value of $\sigma^{(r)}\approx 4.7\times 10^{-18}$\,cm$^2$, while the cross section for direct ICEC amounts to $\sigma^{(d)}\approx 8.9\times 10^{-26}$\,cm$^2$. The widths of the resonance peak and of the associated Fano structure are much more narrow in this case.

The enormous enhancement shown in Fig.~\ref{fig:energy} relies on the assumption that the incident electron energy meets the resonance condition exactly. However, in an experiment the incident electron beam will contain a distribution of energies around the resonant value with a certain width $\delta\varepsilon$. The latter will typically be $\delta\varepsilon\gg\Gamma$, so that only a small fraction $\sim\Gamma/\delta\varepsilon\ll 1$ of all electrons can effectively contribute to the resonant process of three-center rICEC. In contrast, the direct ICEC between atoms $A$ and $C$ is nonresonant; its cross section therefore depends only weakly on $\varepsilon_q$ and remains practically unchanged when an electron beam with narrow energy distribution is applied. Assuming as an example $\delta\varepsilon\sim 10^{-2}$\,eV, $\Gamma\approx\Gamma_{\rm rad}^{(B)}\sim 10^{-6}$\,eV and, as before, $\omega_{ge}\approx 20$\,eV, $R\approx 10$\,\AA, the ratio of the energy-averaged cross sections becomes [see Eq.~\eqref{CS-2}]
$$ \frac{\langle\sigma^{(r)}\rangle}{\langle\sigma^{(d)}\rangle} \approx \frac{9\Gamma}{\delta\varepsilon} \left(\frac{cR_{AC}}{\omega_{ge} R_{BA}R_{BC}}\right)^{\!6} \sim 10^3\ .$$
Thus, the enhancement of three-center rICEC over direct ICEC is still very large. For $R_{BA}\approx R_{BC}\approx \frac{1}{2}R_{AC}$ (see Fig.~\ref{fig:coordinates}), the ratio even reaches $\sim 10^5$.

\subsection{Geometry dependence}
Next we analyze how the three-center rICEC cross section depends on the geometry of the triatomic system that is defined by the internuclear vectore ${\bf R}_{BA}$ and ${\bf R}_{BC}$. The energy of the incident electron is assumend to be resonant, $\varepsilon_q = \varepsilon_0 + \omega_{ge}\approx 7.62$\,eV.

\begin{figure}[b]
\begin{center}
\includegraphics[width=0.35\textwidth]{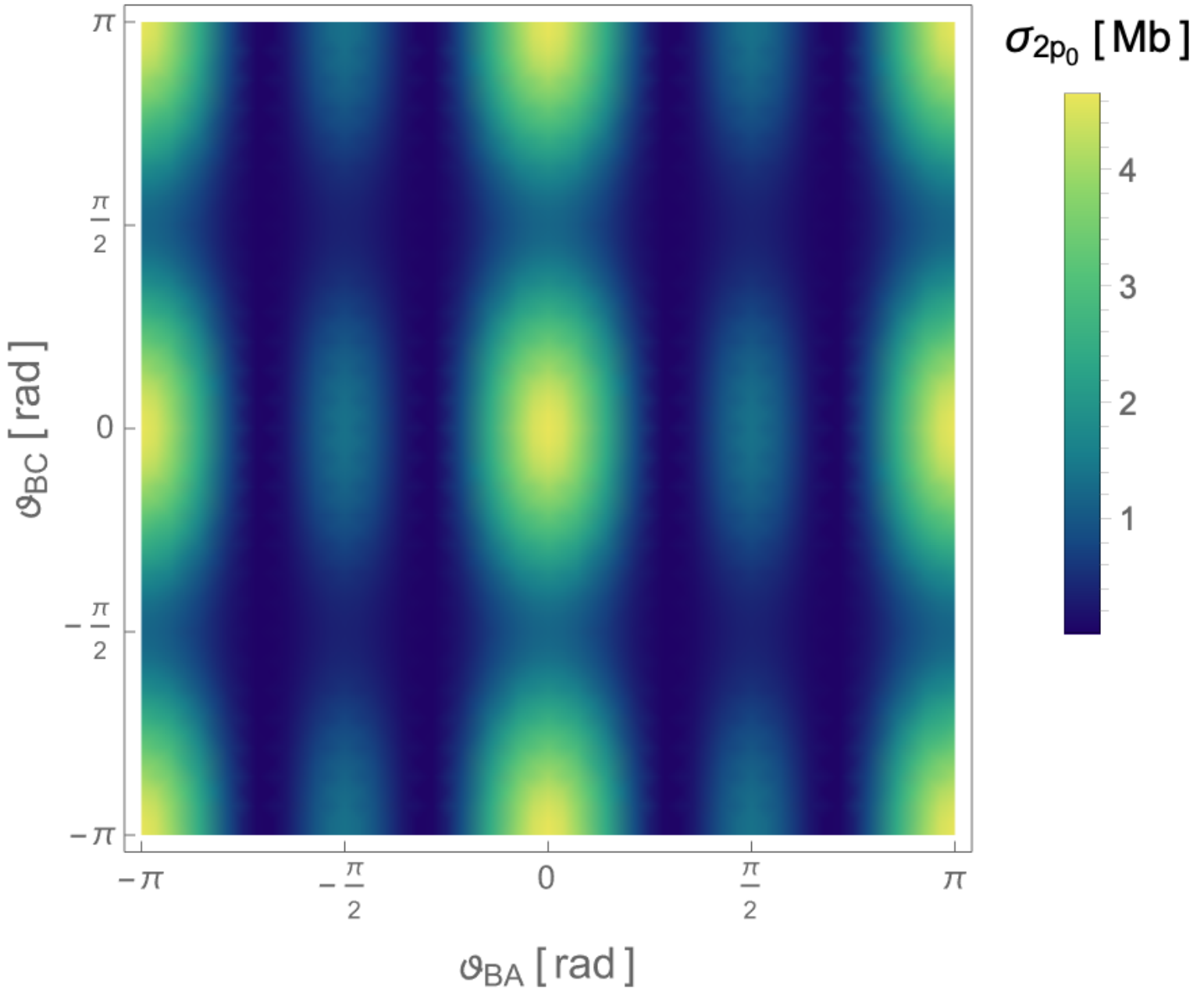}
\includegraphics[width=0.35\textwidth]{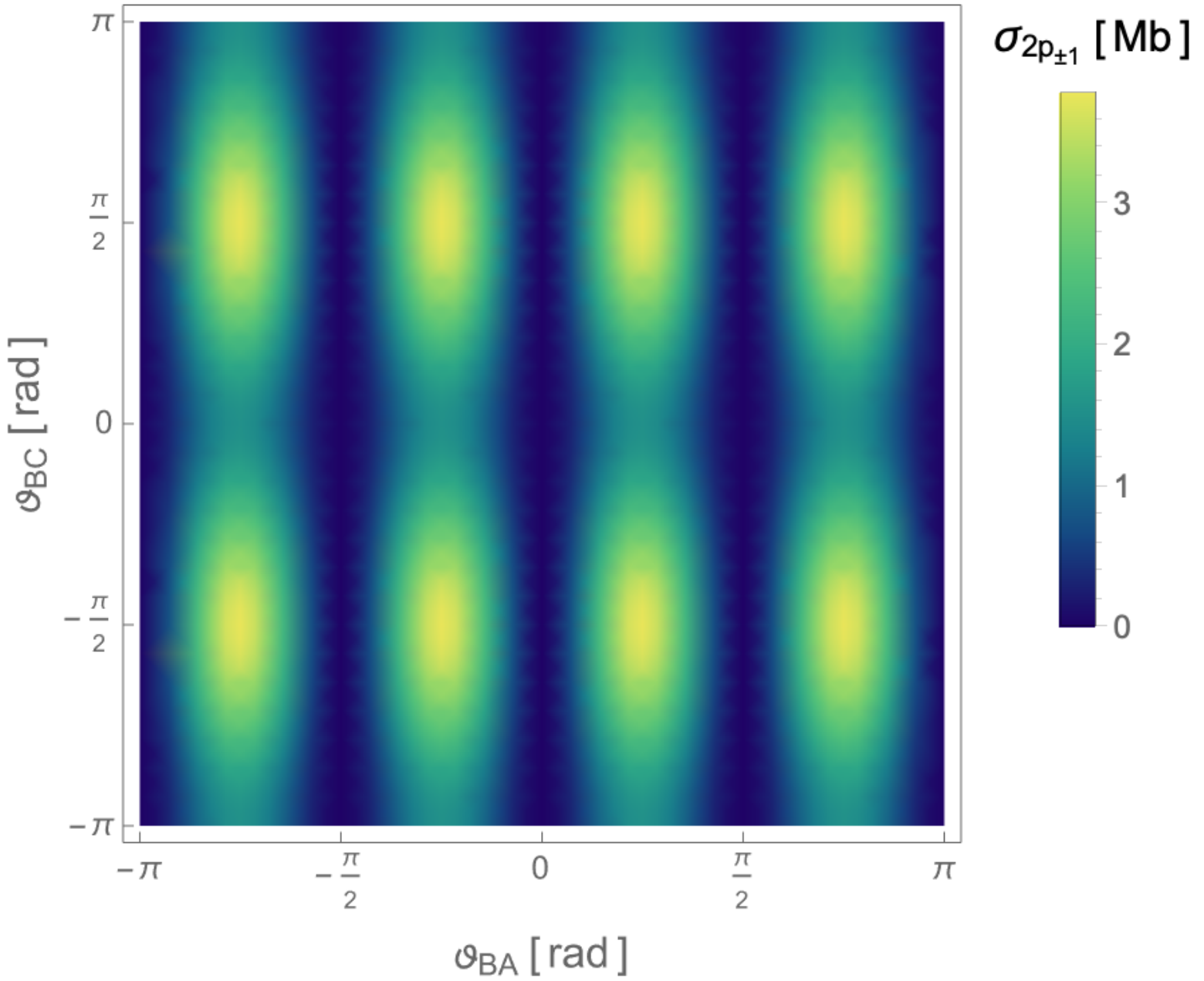}
\includegraphics[width=0.35\textwidth]{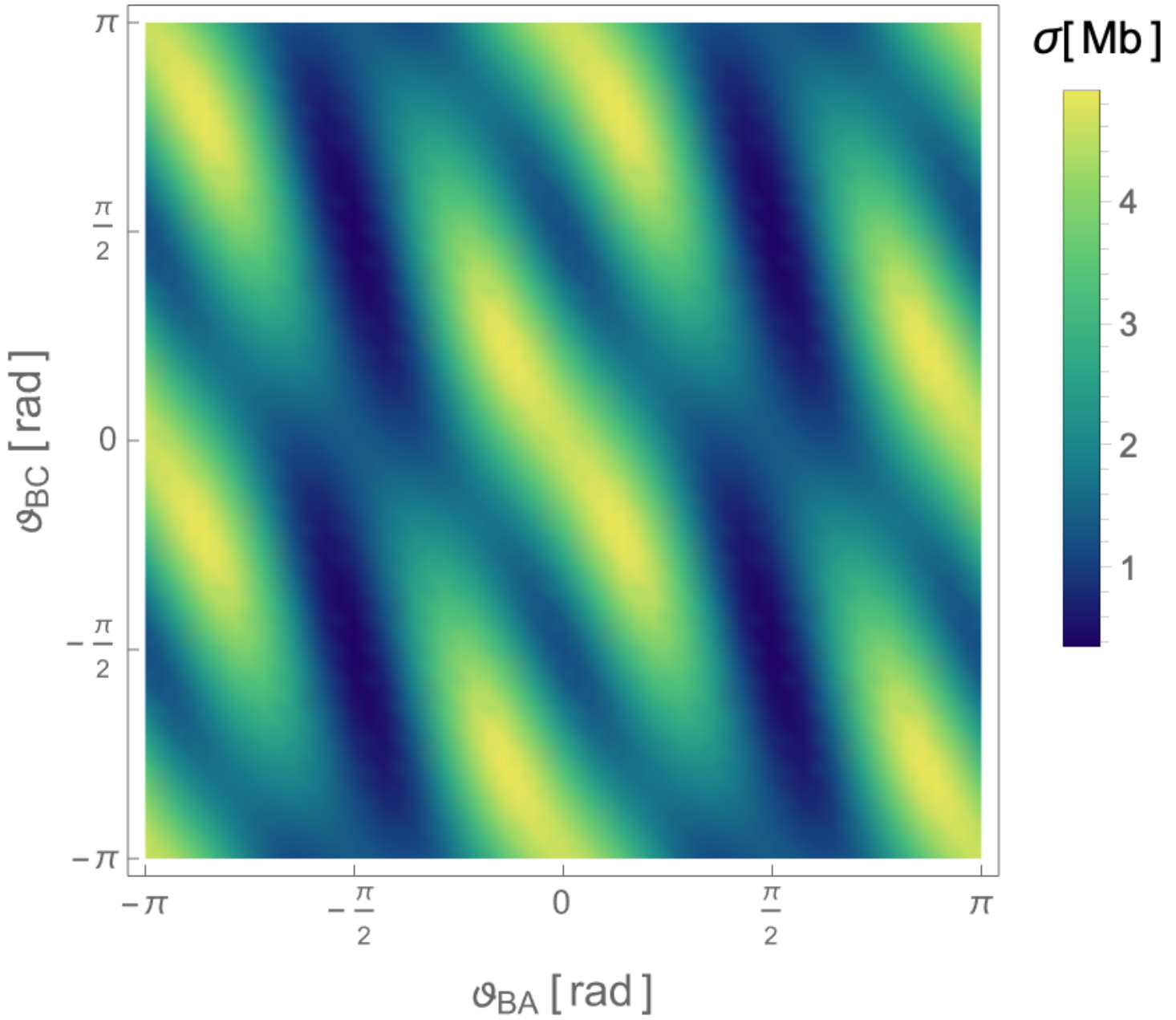}
\caption{Cross sections of three-center rICEC in a H-He-Li system, located in the $xz$ plane. Shown is the dependence on the relative interatomic orientations that are parametrized by the polar angles $\vartheta_{BA}$ and $\vartheta_{BC}$ of the internuclear separation vectors ${\bf R}_{BA}$ and ${\bf R}_{BC}$, respectively. The distances are kept constant, $R_{BA}=R_{BC}=25$\,a.u., and the incident electron energy satisfies the resonance condition. The top panel (a) [middle panel (b)] shows the partial cross section $\sigma_{2p_0}^{(r)}$ [$\sigma_{2p_{\pm 1}}^{(r)}$] and the bottom panel (c) the total cross section $\sigma^{(r)}$.}
\label{fig:geo-25}
\end{center}
\end{figure}

Figure~\ref{fig:geo-25} illustrates the dependence of the three-center rICEC process on the polar angles $\vartheta_{BA}=\arccos(R_{BA,z}/R_{BA})$ and $\vartheta_{BC}=\arccos(R_{BC,z}/R_{BC})$, when the atoms are located in the $xz$ plane.\footnote{This configuration represents the most interesting case. In the complementary setting where the atoms lie in the plane perpendicular to the incident electron momentum, $\sigma_{2p_0}^{(r)}$ attains a constant, angle-independent value and $\sigma_{2p_{\pm 1}}^{(r)}=0$.} The internuclear distances are taken as $R_{BA}=R_{BC}=25$\,a.u. In this situation the radiative width $\Gamma_{\rm rad}^{(B)}$ is much larger than the ICD widths. Panel (a) shows the contribution to the cross section stemming from excitation of the $1s2p_0$ state in helium and panel (b) the corresponding contribution from the $1s2p_{+1}$ (or $1s2p_{-1}$) state. In the total transition amplitude, however, the partial contributions from the three excited states have to be summed coherently, which leads to quantum interference effects [see the comment below Eq.~\eqref{S1}]. The geometry dependence of the total cross section of three-center rICEC, including these interference effects, is displayed in panel (c).

The angular structures in Fig.~\ref{fig:geo-25}\,(a) and (b) can be understood as follows. When the corresponding terms in the transition amplitude are squared separately (i.e.~incoherently), one obtains the associated partial contributions to the cross section, whose dependence on the internuclear angles is given by
\begin{eqnarray*}
\sigma_{2p_0}^{(r)}\!&\propto&\! \big(3\cos^2\!\vartheta_{BA}-1\big)\big(3\cos^2\!\vartheta_{BC}+1\big)\,, \\
\sigma_{2p_{\pm 1}}^{(r)} \!&\propto&\! \big( 3\sin\vartheta_{BA}\cos\vartheta_{BA}\big)^2
\big(3\sin^2\!\vartheta_{BC}+2\big)\,.
\end{eqnarray*} 
The asymmetry in the angles $\vartheta_{BA}$ and $\vartheta_{BC}$ arises from the fact that the momentum ${\bf q}$ of the incident electron, that is captured by atom $A$, is fixed along the $z$ axis, whereas the momentum ${\bf p}$ of the electron ejected from atom $C$ is integrated over emission angles.

According to the first of the above equations, the partial cross section $\sigma_{2p_0}^{(r)}$ is maximized for $\vartheta_{BA},\vartheta_{BC}\in\{0,\pm\pi\}$, in agreement with Fig.~\ref{fig:geo-25}\,(a). Conversely, this contribution vanishes for angles $\vartheta_{BA}$ fulfilling the relation $\cos^2\!\vartheta_{BA}=1/3$. These special values are generally known in the literature as 'magic angles'. They arise typically in processes that rely on dipole-dipole interactions, such as ICD \cite{magic1}, two-center photoionization \cite{magic2} or nuclear magnetic resonance spectroscopy \cite{magic3}. 

Conversely, as the second of the above equations and Fig.~\ref{fig:geo-25}\,(b) show, the partial contribution from excitation of a $1s2p_{\pm 1}$ state is maximized for $\vartheta_{BA}\in\{\pm\frac{\pi}{4},\pm\frac{3\pi}{4}\}$, $\vartheta_{BC}\in\{\pm\frac{\pi}{2}\}$ and minimized for $\vartheta_{BA}\in\{0,\pm\frac{\pi}{2}\pm\pi\}$. 

The partial cross sections in Fig.~\ref{fig:geo-25}\,(a) and (b) exhibit a clear regularity and two separate mirror symmetries under both $\vartheta_{BA}\to-\vartheta_{BA}$ as well as $\vartheta_{BC}\to-\vartheta_{BC}$. The total cross section in Fig.~\ref{fig:geo-25}\,(c), involving a coherent sum over the contributions from all three excited states, has a less regular appearance and remains symmetric only under the combined transformation $(\vartheta_{BA},\vartheta_{BC})\to-(\vartheta_{BA},\vartheta_{BC})$. This is caused by the quantum interferences that may be either constructive or destructive, depending on the spatial geometry of the triatomic system.

\begin{figure}[b]
\begin{center}
\includegraphics[width=0.4\textwidth]{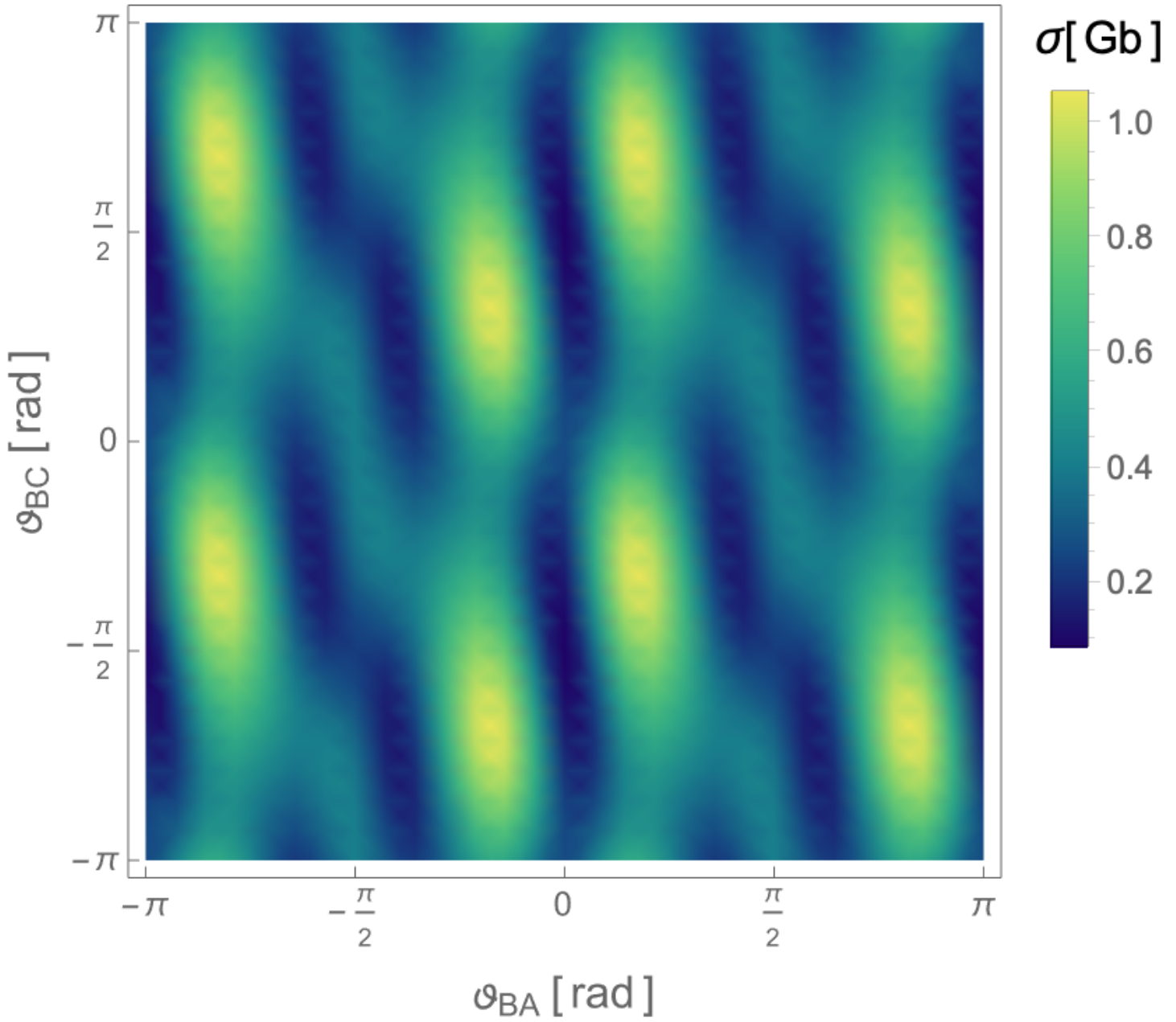}
\caption{Same as Fig.~\ref{fig:geo-25}\,(c) but for $R_{BA}=R_{BC}=10$\,a.u.}
\label{fig:geo-10}
\end{center}
\end{figure}

The results in Fig.~\ref{fig:geo-25} have been obtained for specific values of the  internuclear distances. Such an arrangement of three atomic centers at customized positions and relative orientations could be feasible to realize with quantum dot systems. We note, however, that the shape of the angular structures is robust: it would remain the same if other internuclear separations (including $R_{BA}\ne R_{BC}$) were used, as long as the radiative width largely exceeds the ICD widths and the interference with direct ICEC is negligible. Only the absolute level of the cross sections would change then. Also the application of incident electron energies that are not exactly resonant but lie slightly off the resonance would not alter the angular structures.

The geometry dependence changes considerably at smaller internuclear distances where the ICD widths become comparable to or even exceed the radiative width. In this situation, not only the dipole matrix elements in Eq.~\eqref{S3} depend on ${\bf R}_{BA}$ and ${\bf R}_{BC}$ but also the width $\Gamma$ in the denominator is sensitive to the interatomic orientations. For the different excited states in atom $B$, the angular dependencies are given by $\Gamma_{{\rm ICD},2p_0}^{(BX)}\propto 3\cos^2\!\vartheta_{BX}+1$ and $\Gamma_{{\rm ICD},2p_{\pm 1}}^{(BX)}\propto 3\sin^2\!\vartheta_{BX}+2$, with $X\in\{A,C\}$ labelling the ICD involving atom $A$ or atom $C$, respectively. The 'magic' $\vartheta_{BA}$ angles, where the partial cross sections $\sigma_{2p_0}^{(r)}$ or $\sigma_{2p_{\pm 1}}^{(r)}$ vanish, thus remain unaltered, but the dependence on $\vartheta_{BC}$ turns out to be somewhat weaker. The total cross section, including quantum interferences, attains the form shown in Fig.~\ref{fig:geo-10} which differs considerably from the outcome in Fig.~\ref{fig:geo-25}\,(c). In particular, the maximum at the center of the plot ($\vartheta_{BA}=\vartheta_{BC}=0$) has disappeared and the overall structure has become more complex. 

\medskip

Finally, it is worth mentioning that the process of three-center rICEC in our model system may also proceed in inverse order, according to
$$ e' + {\rm H} + {\rm He} + {\rm Li}^+ \to {\rm H} + {\rm He}^* + {\rm Li} \to {\rm H}^+ + {\rm He} + {\rm Li} + e\ .$$
Here, the resonance energy of the incident electron would be $\approx 15.83$\,eV.

\section{Conclusion}
The process of three-center rICEC has been studied where an incident electron is first captured by an atomic center $A$, with the excess energy being transfered radiationlessly to a neighbouring atom $B$ that is resonantly excited. In a second step, the excitation energy is again transfered radiationlessly to yet another neighbouring atom $C$, leading to its ionization via ICD. We have shown that three-center rICEC, due to it resonant nature, can strongly dominate over the direct (nonresonant) ICEC between atoms $A$ and $C$ and also over the single-center process of radiative recombination with atom $A$. The resonant enhancement can be so strong that the dominance of three-center rICEC may persist even after averaging over the energy distribution of an incident electron beam. When the distance between atoms $B$ and $C$ is not too large, three-center rICEC also dominates over the resonant diatomic process of 2CDR, where atom $B$ deexcites via photoemission. Thus, under suitable conditions, three-center rICEC may indeed constitute the strongest channel for electron capture into center $A$. By considering a simple triatomic model system composed of H, He and Li, we have moreover analyzed the rather complex dependence of three-center rICEC on the relative interatomic orientations. 

Three-center rICEC could in principle be observed in protonated heteroatomic noble-gas dimers, such as 
He-H$^+$-Ng with ${\rm Ng} \in\{ {\rm Ne}, {\rm Ar}, {\rm Kr}, {\rm Xe}\}$ \cite{protonated1,protonated2}. In analogy to our model system, the incident electron would be captured by the proton, with resonant excitation of the He atom, which afterwards could transfer the energy to the other noble-gas atom that is ionized. We should note, however, that the internuclear distances $\sim 3$-4 a.u. in such protonated noble-gas dimers are not very large \cite{protonated1,protonated2}, so that the theoretical approach developed in the present paper is not perfectly suitable for them. In addition, one would need to take the nuclear motion in the molecules into account \cite{2CPI-mol}. Alternatively, systems of three quantum dots could be considered. They offer the advantage that the intersite distances can be fixed and controlled \cite{ICEC-dots1,ICEC-dots2,ICEC-dots3}.

\section*{Acknowledgement} 
Useful input by A. G\"orlitz is gratefully acknowledged.



\begin{thebibliography}{33}

\bibitem{Review-Recomb1}
Hahn Y 1997 {\it Rep. Prog. Phys.} {\bf 60} 691

\bibitem{Review-Recomb2}
Beiersdorfer P 2003 {\it Annu. Rev. Astron. Astrophys.} {\bf 41} 343

\bibitem{Review-Recomb3}
M\"uller A 2008 {\it Adv. At. Mol. Opt. Phys.} {\bf 55} 293

\bibitem{ICD-Review1}
Hergenhahn U 2011 {\it J. Electron Spectrosc. Relat. Phenom.} {\bf 184} 78

\bibitem{ICD-Review2} 
Jahnke T 2015 {\it J. Phys. B} {\bf 48} 082001

\bibitem{ICD-Review3}
Jahnke T, Hergenhahn U, Winter B, D\"orner R, Fr\"uhling U, Demekhin P~V, Gokhberg K, 
Cederbaum L~S, Ehresmann A, Knie A and Dreuw A 2020 {\it Chem. Rev.} {\bf 120} 20

\bibitem{2CDR-PRL}
M\"uller C, Voitkiv A~B, Crespo L\'opez-Urrutia J~R and Harman Z 2010 
{Phys. Rev. Lett.} {\bf 104} 233202

\bibitem{2CDR-PRA2010}
Voitkiv A~B and Najjari B 2010 {\it Phys. Rev.} A {\bf 82} 052708

\bibitem{2CDR-PRA2018}
Eckey A, Jacob A, Voitkiv A~B and M\"uller C 2018
{\it Phys. Rev.} A {\bf 98} 012710

\bibitem{2CDR-coll}
Jacob A, M\"uller C and Voitkiv A~B 2019
{\it Phys. Rev.} A {\bf 100} 012706

\bibitem{2CDR-coll-Hbar}
Jacob A, Zhang S~F, M\"uller C, Ma X and Voitkiv A~B 2020
{\it Phys. Rev. Res.} {\bf 2} 013105

\bibitem{2CPI} Najjari B, Voitkiv A~B and M\"{u}ller C 2010 
{\it Phys. Rev. Lett.} {\bf 105} 153002

\bibitem{2CPIexp}
Trinter F \textit{et al.} 2013 {\it Phys. Rev. Lett.} \textbf{111} 233004\\
Mhamdi A {\it et al.} 2018 {\it Phys. Rev.} A {\bf 97} 053407

\bibitem{Hergenhahn} Hans A, Schmidt P, Ozga C, Richter C, Otto H, Holzapfel X, Hartmann G,
Ehresmann A, Hergenhahn U and Knie A 2019 {\it J. Phys. Chem. Lett.} {\bf 10} 1078

\bibitem{ICEC-JPB}
Gokhberg K and Cederbaum L S 2009 {\it J. Phys. B: At. Mol. Opt. Phys.} {\bf 42} 231001 (FTC) 

\bibitem{ICEC-PRA}
Gokhberg K and Cederbaum L S 2010 {\it Phys. Rev.} A {\bf 82} 052707

\bibitem{ICEC-Sisourat}
Sisourat N, Miteva T, Gorfinkiel J~D, Gokhberg K and Cederbaum L~S 2018
{\it Phys. Rev.} A {\bf 98} 020701(R)

\bibitem{Matthew} Matthew J and Komninos Y 1975 {\it Surf. Sci.} {\bf 53} 716

\bibitem{ICD} Cederbaum L~S, Zobeley J and Tarantelli F 1997 
{\it Phys. Rev. Lett.} \textbf{79} 4778

\bibitem{ICEC-coll}
Jacob A, M\"uller C and Voitkiv A B 2019
{\it J. Phys. B: At. Mol. Opt. Phys.} {\bf 52} 225201

\bibitem{ICEC-dots1}
Pont F~M, Bande A and Cederbaum L~S 2013 {\it Phys. Rev.} B {\bf 88} 241304(R)

\bibitem{ICEC-dots2}
Pont F~M, Bande A and Cederbaum L~S 2016 {\it J. Phys.: Cond. Matter} {\bf 28} 075301

\bibitem{ICEC-dots3}
Molle A, Berikaa E, Pont F~M and Bande A 2019 {\it J. Chem. Phys.} {\bf 150} 224105

\bibitem{ICEC-water}
Molle A, Dubois A, Gorfinkiel J~D, Cederbaum L~S and Sisourat N 2021
{\it Phys. Rev.} A {\bf 103} 012808 \\
Molle A, Dubois A, Gorfinkiel J~D, Cederbaum L~S and Sisourat N 2021
{\it Phys. Rev.} A {\bf 104} 022818

\bibitem{res-ICEC1}
Bande A, Pont F~M, Gokhberg K and Cederbaum L~S 2015 
{\it EPJ Web of Conferences} {\bf 84} 07002

\bibitem{res-ICEC2}
Pont F~M, Molle A, Berikaa E, Bubeck S and Bande A 2019 
{\it J. Phys.: Cond. Matter} {\bf 32} 065302

\bibitem{super-ICD}
Miteva T, Kazandjian S, Koloren{\v c} P, Votavov\'a P and Sisourat N 2017 
{\it Phys. Rev. Lett.} {\bf 119} 083403

\bibitem{Exp1}
Yan S {\it et al.} 2013 {\it Phys. Rev. A} {\bf 88} 042712 (2013)

\bibitem{Exp2}
Pfl\"uger T, Ren X and Dorn A 2015 {\it Phys. Rev. A} {\bf 91} 052701

\bibitem{Exp3}
Yan S {\it et al.} 2018 {\it Phys. Rev.} A {\bf 97} 010701(R)

\bibitem{Exp4}
Ren X, Al Maalouf E~J, Dorn A and Denifl S 2016 {\it Nature Commun.} {\bf 7} 11093

\bibitem{ICD-formula}
Gr\"ull F, Voitkiv A B and M\"uller C 2019 {\it Phys. Rev.} A {\bf 100} 032702

\bibitem{LL}
Landau L D and Lifshitz E M 1965 {\it Quantum Mechanics} (Pergamon, Oxford)

\bibitem{NIST} Atomic spectra database of the National Institute of Standards 
and Technology (NIST), available at https://www.nist.gov/pml/atomic-spectra-database


\bibitem{magic1}
Cederbaum L S and Kuleff A I 2021 {\it Nature Commun.} {\bf 12} 4083

\bibitem{magic2} Here, the magic angle is encoded in an 'effective polarization angle', see 
Eckey A, Voitkiv A~B and M\"uller C 2020 {\it J. Phys. B: At. Mol. Opt. Phys.} {\bf 53} 055001

\bibitem{magic3}
Mehring M and Waugh J S 1972 {\it Phys. Rev.} B {\bf 5} 3459\\
Andrew E R, Richards R E and Packer K J 1981 {\it Phil. Trans. Royal Soc. A} {\bf 299} 505

\bibitem{protonated1}
Borocci S, Grandinetti F and Sanna N 2021 {\it Molecules} {\bf 26} 1305

\bibitem{protonated2}
Tan J~A and Kuo J~L 2022 {\it Molecules} {\bf 27} 3198

\bibitem{2CPI-mol}
Gr\"ull F, Voitkiv A~B and M\"uller C 2022 {\it J Phys B: At. Mol. Opt. Phys.} {\bf 55} 245101


\end{thebibliography}
\end{document}